\documentclass[aps,pra,nofootinbib,notitlepage,twocolumn,superscriptaddress]{revtex4-1}
\usepackage{graphicx}
\usepackage{multirow}
\usepackage{amsmath}
\usepackage{amssymb}
\usepackage{amsthm}
\usepackage{eucal}
\usepackage{bm}
\usepackage{upgreek}
\usepackage{framed}
\usepackage{mathrsfs}
\usepackage{latexsym}
\usepackage{epstopdf}
\usepackage{subfigure}
\usepackage{color}
\newcommand{\be}{\begin{equation}}
\newcommand{\ee}{\end{equation}}
\newcommand{\bea}{\begin{eqnarray}}
\newcommand{\eea}{\end{eqnarray}}
\newcommand{\ba}{\begin{array}}
	\newcommand{\ea}{\end{array}}
\newcommand{\pala}{\partial_\lambda}

\begin{document}	
	
	\title{Quantum Critical Environment Assisted Quantum Magnetometer}
	
	\author{Noufal Jaseem}
	\email{noufaljaseemp@iisertvm.ac.in}
	\affiliation{School of Physics, Indian Institute of Science Education and Research Thiruvananthapuram, India 695016}
	
	\author{S.~Omkar}
	\affiliation{Centre for Macroscopic Quantum Control, Department of Physics and Astronomy, Seoul National University, Seoul, South Korea}
	
	\author{Anil Shaji}
	\affiliation{School of Physics, Indian Institute of Science Education and Research Thiruvananthapuram, India 695016}

	\begin{abstract}
		A central qubit coupled to an Ising ring of $N$ qubits, operating close to a critical point is investigated as a potential precision quantum magnetometer for estimating an applied transverse magnetic field. We compute the Quantum Fisher information for the central, probe qubit with the Ising chain initialized in its ground state or in a thermal state. The non-unitary evolution of the central qubit due to its interaction with the surrounding Ising ring enhances the accuracy of the magnetic field measurement. Near the critical point of the ring,  Heisenberg-like scaling of the precision in estimating the magnetic field is obtained when the ring is initialized in its ground state. However, for finite temperatures, the Heisenberg scaling is limited to lower ranges of $N$ values. 
	\end{abstract}
	\pacs{03.65.Ta, 06.20.-f, 05.70.Jk, 05.30.Rt}
	\keywords{quantum metrology, quantum Fisher information, quantum critical system, Loschmidt echo}
	\maketitle
	
	\section{Introduction}
	
	In making precision measurements, what matters is how much the state of the measuring device changes when the measured quantity changes a bit. If the state of the device changes substantially with small changes in the value of the measured quantity, then one obtains a sensitive measurement with small uncertainties. On the other hand, if the state of the device hardly changes in response to  changes in the measured quantity then the uncertainties associated with the measurement are large. In quantum-limited measurements, the measuring device is the quantum state of a probe system. In the following we consider single parameter estimation. The rate of change of the state of the probe in response to changes in the value of the measured parameter determines the measurement precision. This idea is captured by the quantum Cramer-Rao bound~\cite{helstrom76a,holevo82a,braunstein94a,braunstein96a} that puts the theoretical lower bound on the achievable measurement precision in a quantum metrology setup as
	\[ (\delta \lambda)^2 \geq \frac{1}{m {\mathcal F}}, \]
	where $\lambda$ is the measured parameter, $m$ is the number of repetitions of the measurement and ${\mathcal F}$ is the quantum Fisher information (QFI)~\cite{fisher1925,Hig+07,GLM11}  
	
	Typically the quantum probe is assumed to be made up of $N$ sub-units. The number of sub-units is essentially a placeholder for the most significant limiting resource that goes into the measurement process. It may very well be the finite time available for performing the measurement, the maximum energy that can be deposited on a sample without damaging it, the mean photon number in the arms of an interferometer or just the number $N$ of atoms used in a Ramsey type interferometry experiment~\cite{katarzynaTime,Taylor20161review,Caves:1980vh,wineland1992Hscaling,GLM2004,GLM11}. In characterizing metrology schemes, the scaling of the measurement uncertainty with $N$ is an important figure of merit. A classical probe with $N$ sub-units is limited to achieving a scaling of $1/\sqrt{N}$, which is called the shot noise limit. A quantum probe can improve the scaling to the Heisenberg-limited one of $1/N$ by marshaling quantum resources like entanglement and coherence~\cite{bondurantHscaling,yurkeHscaling,dowling1998Hscaling,BollingerHscaling,wineland1992Hscaling,demkowicz2015chapter,dowling2008,GLM2004}.
	
	Pursuing the idea of a rapid change in the state of the quantum probe with respect to the measured parameter further we recognize that a scenario in which the state of a quantum many-body system undergoes a rapid, parameter dependent change is across a phase transition. It is therefore natural to ask whether such a phase transition can be leveraged to perform single parameter estimation with enhanced sensitivities. This question was answered in the affirmative in~\cite{Zanardi:2008ih,Invernizzi:2008ez}  where quantum criticality was identified as a potential resource for enhanced quantum limited metrology. In~\cite{Invernizzi:2008ez} an example of the estimation of the coupling parameter of a one-dimensional quantum Ising model was considered in the specific cases of a few spins, as well as in the thermodynamic limit. It was shown that at the critical point Heisenberg-limited scaling of the measurement precision, $\delta \lambda$,  that is of the same order as the length of the Ising chain (inversely proportional to the number $N$ of spins in the chain) is obtained. In~\cite{Zanardi:2008ih}, the example considered is the estimation of a coupling in a fermionic tight-binding (BCS-type) mode.  Going beyond models with nearest neighbor interactions that exhibit phase transitions, a similar analysis was done on the Lipkin-Meshkov-Gleick model in~\cite{Salvatori:2014iy}. When considering such phase transitions~\cite{Zanardi:2008ih,Invernizzi:2008ez,Salvatori:2014iy} the parameter dependent evolution of the probe is considered to be unitary. Non-unitary processes that are considered, if at all, are as noise on the system. 
	
	We consider a generalised Hepp-Coleman~\cite{Bel1975,Hep1972,QSL+06} model where an Ising ring of spin $1/2$ particles, placed in an external magnetic field, is coupled to a central probe spin. The central spin is also a two-level quantum system that we refer to as the probe qubit. We use the probe qubit to measure an unknown magnetic field by bringing the system very close to the critical point of the ring.  We demonstrate that very close to the critical point of the ring, there is a steep increase in the QFI of the probe qubit when it is used to measure the external magnetic field. We study the scaling of the QFI of the decohering probe qubit with the number of spins in the Ising ring coupled to it. Operating the quantum probe close to a phase transition can always be done using an adaptive, iterative scheme. The estimate of the external parameter, which is the magnetic field in this case, is progressively refined at each step and compensatory fields adjusted so as to keep the Ising ring always close to the critical point.
	
	The system we consider has a finite number $N$ of spins in the Ising ring. Strictly speaking, the quantum phase transition for the ring appears in the thermodynamic limit wherein $N \rightarrow \infty$. In the following when we mention critical points and phase transitions, we refer to regimes where the relevant parameter values produce a phase transition in the large $N$ limit. However, as we will see, the behavior of the finite $N$ system is interesting enough around these values to merit closer attention as a tool for performing quantum limited measurements. In fact, the thermodynamic ($N\rightarrow \infty$) limit is not particularly interesting in the the metrology context because, if one allows for infinite resources, then arbitrarily small precision is possible. So for finite $N$, close to the critical point of the system we show that the QFI of the probe qubit is directly proportional to the square of the number of the spins in the Ising chain. 
	
	A system similar to the one we discuss in this work is considered in the context of quantum-limited measurements in~\cite{centralQFI}, where the QFI of a central spin surrounded by a quantum critical spin chain is computed. However, in~\cite{centralQFI} the quantum critical environment furnished by the Ising ring is taken as a source of noise. In this paper,  apart from using the phase transition for metrology, the other key idea we present is that the quantum critical behavior of the environment of the quantum probe leads to rapid, parameter dependent, non-unitary changes in its state and this non-unitary evolution can be leveraged to improve the precision of the measurement. The general formulation of the quantum Cramer-Rao bound~\cite{braunstein94a,braunstein96a} allows one to consider non-unitary parameter dependent evolution as well. Environment-assisted quantum metrology has been studied previously as well~\cite{EnvAssistQM_PRL, EnvAssistQM_PRA} but not in the context of the environment having a phase transition.
	
	Our work is motivated by the fact that detecting very weak magnetic fields with high precision is important to many fields like material science, data storage, biomedical science, navigation and fundamental physics. Particularly in investigating magnetism at the nano and atomic scales in spintronic and magnetism based devices, where the single atom sensors play a significant role. Magnetic field sensors have been developed using a wide variety of approaches. These include superconducting quantum interference devices~\cite{Ben1999}, the Hall effect in semiconductors~\cite{Bud+02}, atomic vapours~\cite{Auz+04,SSR+05},  Bose-€"Einstein condensates~\cite{ven+07,ZW06} and magnetic resonance force microscopy~\cite{MPD+07, Sch+1999}. Single-spin as a probe to sense the magnetic field has also been studied previously. For instance, the single-spin associated with the nitrogen-vacancy (N-V) centers in the diamond that provide long spin-life under normal environmental conditions~\cite{Deg08,TCC+08,Bal+08,Bon+15} have been proposed for magnetometry. 
	
	This article is structured as follows: In section \ref{centralsystemF}, we derive the QFI of the probe qubit. In section \ref{groundFid} we discuss the QFI when the Ising ring is initialized in its ground state. The case where the ring is initialized in a thermal state is investigated in section \ref{thermalFid}. Our findings are summarized in section~\ref{conclude}.
	
	\section{Quantum Fisher Information for a decohering qubit \label{centralsystemF}}
	We label the probe qubit $S$, and it is coupled to an Ising ring $E$ that forms the immediate environment of the probe. Both $S$ and $E$ are placed in the transverse magnetic field aligned along the $x-$axis, the magnitude of which is to be measured. A tunable magnetic field is also assumed to be present that enables the iterative, adaptive scheme that keeps the $SE$ system close to the critical point of $E$ at each step.  The state $|1\rangle$ of the central qubit is coupled to $E$, and the Hamiltonian for the $SE$ system is~\cite{LEcritical},
	\be
	H(\lambda,\delta)= -\sum_j\left(\sigma^z_j\sigma^z_{j+1}+\lambda\sigma^x_j+ \delta|1\rangle\langle 1|\sigma^x_j\right),
	\label{eq:hamil}
	\ee  
	Here $\lambda$ is the magnitude of the transverse magnetic field which is to be estimated, $\delta$ is the coupling strength between the qubit and the spin chain that is assumed to be small enough to be treated perturbatively, and $\sigma^{x,y,z}_j$ are Pauli operators of the $j^{th}$ spin. The lattice spacing for the Ising ring is taken to be unity.  We have assumed that there is no free evolution of $S$ due to the magnetic field $\lambda$ so as to focus on the non-unitary evolution of the probe qubit and its contribution to the measurement sensitivity which is the focus of this paper. 
	
	In the absence of the central, probe qubit, the transverse magnetic field becomes equal in strength to the nearest neighbour coupling between the Ising spins when $\lambda=1$ and this corresponds to the critical point of the Ising ring in the large $N$ limit. When we add in the effect of the coupling to the probe qubit, we expect the critical point to be shifted slightly so that $\lambda + \delta \simeq 1$. Our interest lies in the sensitivity of the probe qubit to small variations in the external magnetic field ($\lambda$) to be measured, particularly in the vicinity of the critical point near $\lambda = 1-\delta$. The sensitivity is quantified in terms of the QFI of the state of the probe as~\cite{braunstein94a}
	\begin{equation}\label{qfi}
	\mathcal{F}={\rm tr}(\mathcal{R}_{\lambda}^2\rho),
	\end{equation}
	where the symmetric logarithmic derivative operator is given by,	
	\[\mathcal{R}_{\lambda}=\sum_{\{j,k|p_j+p_k\neq 0\}} 2\frac{\langle j|\partial_\lambda \rho|k\rangle}{p_j+p_k} \;|j\rangle\langle k|,\]
	in the basis in which the density matrix of the probe is diagonal, $\rho=\sum_j p_j |j\rangle \langle j|$, and $\partial_\lambda$ is the derivative of $\rho$ with respect to $\lambda$.  
	
	To compute the QFI of the probe qubit, we start from its density matrix in the Bloch representation given by, 
	\be \label{rhoqubit} \rho=\frac{1}{2} (\openone+\vec{a}(\lambda) \cdot \vec{\sigma}),\ee
	where $\vec{a}(\lambda)=\{a_x,a_y,a_z\}^T$ is the Bloch vector which describes the probe state and which, in turn, depends on the parameter $\lambda$. In the expression above, $\vec{\sigma}=\{\sigma_x,\sigma_y, \sigma_z\}^T$, where $\sigma_i$'s are Pauli operators on the probe qubit. The eigenvalues and eigenvectors of $\rho$ respectively are 
	\[p_\pm=\frac{1}{2}(1\pm a); \qquad |p_\pm\rangle=\sqrt{\frac{a\mp a_z}{2a}} \bigg\{\frac{a_z\pm a}{a_x+i a_y},1 \bigg\}^T, \]
	where $a=||\vec{a}||$. Using the eigenvalues and eigenvectors of $\rho$ as well as $\partial_{\lambda}\rho=\frac{1}{2}(\partial_\lambda\vec{a}) \cdot \vec{\sigma}$  in Eq.~(\ref{qfi}) we can write the QFI as
	\begin{equation} \label{FIqubit}
	\mathcal{F}=\frac{(\vec{a} \cdot \partial_\lambda\vec{a})^2}{1-a^2}+(\partial_\lambda\vec{a})^2.
	\end{equation}
	
	The interaction with the Ising chain gives rise to essentially dephasing dynamics of the probe qubit. Assuming that probe qubit is initially in a pure state given by  $|\psi\rangle = \cos \theta |0\rangle+\sin \theta e^{i\omega} |1\rangle$ in the computational basis (eigenbasis of $\sigma_{z}$) with the ground state labeled as $|0\rangle$ and the exited state as $|1\rangle$. Let the initial state of the Ising ring to be $|\phi(0)\rangle$.  The combined state evolves under the joint Hamiltonian, $H(\lambda,\delta)$ described in Eq.~(\ref{eq:hamil}) and at time $t$ becomes   
	\begin{equation}
	\label{eq:psit}
	| \Psi(t)\rangle=     \cos\theta | 0\rangle  \arrowvert\phi_0(t)\rangle + e^{i\omega}  \sin \theta | 1\rangle  \arrowvert \phi_1(t)\rangle
	\end{equation}
	where 
	\[ |\phi_0(t)\rangle = e^{-i H_0 t} |\phi(0)\rangle \;\;  {\rm and} \; \; |\phi_1(t)\rangle = e^{-i H_1 t} |\phi(0)\rangle \]
	with 
	\[ H_0=-\sum_j\left(\sigma^z_j\sigma^z_{j+1}+\lambda\sigma^x_j\right) \;\;  {\rm and}  \;\; H_1=H_0-\delta\sum_j \sigma^x_j. \]
	The states $|\phi_0(t)\rangle$ and $|\phi_1(t)\rangle$ of the ring are not identical because the coupling of the excited state of probe qubit to the Ising ring. Tracing out the Ising ring from the final state in Eq.~(\ref{eq:psit}) we obtain the time evolved state of the probe qubit as
	\begin{equation}
	\rho_s(t)= 
	\left(\begin{array}{cc}
	\cos^2\theta & \sqrt{\mathcal{L}^g(\lambda, t)}  \frac{\sin 2\theta \;e^{i \omega}}{2}  \\
	\sqrt{\mathcal{L}^g(\lambda, t)} \frac{\sin 2\theta \;e^{-i \omega}}{2}  & \sin^2\theta
	\end{array}\right),
	\label{decoheringstate}
	\end{equation}  
	where $\mathcal{L}^g(\lambda, t)$ is a decoherence factor related to the Fidelity or Loschmidt echo~\cite{ullmanfid,lePRL1,lePRL2,QCriticalSys2} given by,
	\begin{equation}
	\mathcal{L}^g(\lambda, t)=  |\langle\phi_0(t)|\phi_1(t)\rangle|^2.
	\label{eq:lecho0}
	\end{equation}
	We discuss $\mathcal{L}^g(\lambda, t)$ in detail in the next two sections since it represents the parameter dependent, non-unitary evolution of the probe qubit which can be used to enhance the sensitivity of the parameter estimation. Using Eqs.~(\ref{qfi}), (\ref{rhoqubit}), (\ref{FIqubit}) and (\ref{decoheringstate}),  we obtain the following expression for the QFI of the probe qubit:,
	\begin{equation}
	\mathcal{F} = \frac{\left(\partial_\lambda \mathcal{L}\right)^2 }{4 \mathcal{L}\left(1- \mathcal{L}\right)} \sin^2 2\theta.
	\label{Fqubit}
	\end{equation}
	From Eq.~(\ref{Fqubit}), we can see that the probe state that maximises the QFI has $\theta = \pi/4$ and therefore has the form
	\[ |\psi\rangle_{\rm in} = ( |0\rangle+ e^{i\omega} |1\rangle)/\sqrt{2}. \] 
	
	\section{Ising chain initialized in the ground state \label{groundFid}}
	
	We initialise the probe qubit in the pure state $|\psi\rangle_{\rm in} = ( |0\rangle+ e^{i\omega} |1\rangle)/\sqrt{2}$ and take $|\phi(0)\rangle$ to be the ground state of the Ising ring. Following~\cite{QSL+06} we obtain, 
	\begin{equation}
	\mathcal{L}^g(\lambda, t)=\prod_{k>0}\left[1-\sin^2(2\alpha_k)\sin^2(\epsilon_k^1 t)\right],
	\label{eq:lecho1a}
	\end{equation}
	with $k=2n\pi/N$ and $n=1,2, \ldots, N/2$ assuming that $N$ is even. In the equation above,
	\[ \epsilon_k^1 =  2\sqrt{1+(\lambda + \delta)^2-2(\lambda + \delta) \cos(k)} \]
	and $\alpha_k = (\theta^0_k - \theta^1_k)/2$ where
	\[ \theta_k^1=\tan^{-1}\left[\frac{\sin(k)}{\lambda + \delta- \cos (k)}  \right], \quad \theta_k^0=\tan^{-1}\left[\frac{\sin(k)}{\lambda - \cos (k)}  \right]. \]
	An outline of the derivation of Eq.~(\ref{eq:lecho1a}) is given in Appendix~\ref{App_thermalLE}. 
	
	In  the expression we obtained for ${\mathcal L}$, the angles $\alpha_{k}$ and the energies $\epsilon_{k}^{1}$ depend on $\lambda$, and so we have,
	\begin{eqnarray}
	\frac{\pala\mathcal{L}^g}{\mathcal{L}^g} &=& -\sum_k  \bigg[  \frac{2 \sin(2\alpha_k)\sin^2(\epsilon_k^1 t) \pala (\sin(2 \alpha_k))}{1-\sin^2(2\alpha_k)\sin^2(\epsilon_k^1 t)}  \nonumber \\
	&& \qquad \qquad + \frac{t \sin^2(2\alpha_k)\sin(2\epsilon_k^1 t)\; \pala\epsilon_k^1 }{1-\sin^2(2\alpha_k)\sin^2(\epsilon_k^1 t)}  \bigg],
	\end{eqnarray}
	where
	\bea
	\sin(2 \alpha_k) & = & \frac{4 \delta \sin(k)}{\epsilon^0_k \epsilon^1_k}, \nonumber \\
	\pala (\sin(2\alpha_k)) & = &  -4\delta \sin(k) \bigg( \frac{\pala \epsilon^0_k}{(\epsilon^0_k)^2 \epsilon^1_k} + \frac{\pala \epsilon^1_k}{\epsilon^0_k (\epsilon^1_k)^2} \bigg), \nonumber \\
	\epsilon^0_k & = &  2\sqrt{1+\lambda^2-2\lambda \cos(k)}, \nonumber \\
	\pala\epsilon_k^0 &=& \frac{4 \left[ \lambda -\cos\left(k\right)\right]}{\epsilon_k^0}, \nonumber \\
	\pala\epsilon_k^1 &=& \frac{4 \left[\delta +\lambda -\cos\left(k\right)\right]}{\epsilon_k^1}.
	\eea
	Substituting the above results into Eq.~(\ref{Fqubit}), the QFI of the probe qubit for the case in which the Ising ring is initialised in its ground state is obtained as,
	\begin{eqnarray} 
	\label{eq:spin0}
	\mathcal{F}^g(\lambda, t)  & = &  \frac{ {\mathcal{L}}^g }{4 \left[1- {\mathcal{L}}^g\right]} \bigg( \frac{\pala\mathcal{L}^g}{\mathcal{L}^g} \bigg)^2.
	\end{eqnarray}       
	
\begin{figure}[!htb]
		\resizebox{8 cm}{5 cm}{\includegraphics{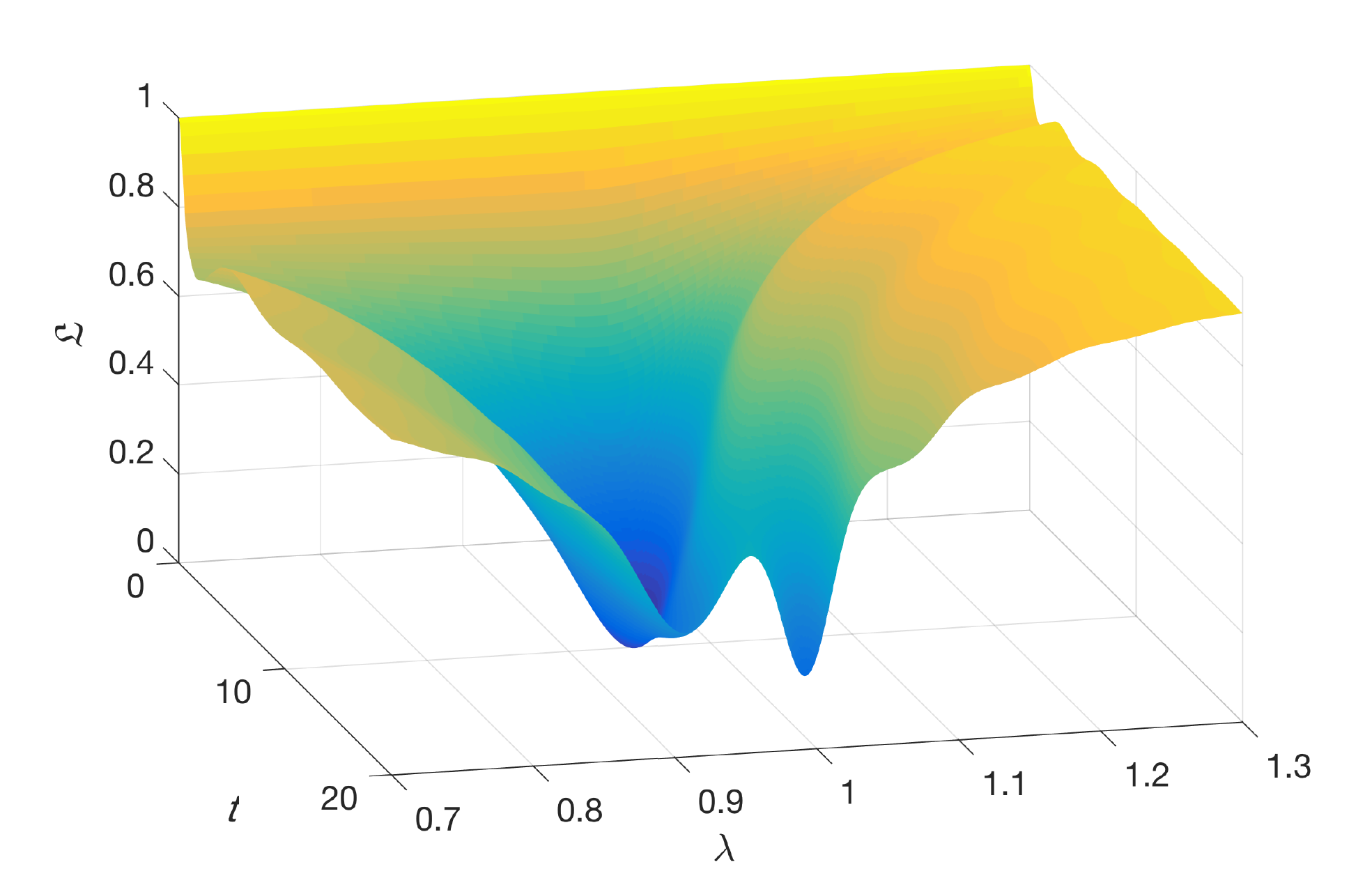}}
		\caption{The decoherence factor ${\mathcal{L}}^g(\lambda, t)$ is plotted against $t$ and $\lambda$ for $N = 100$ and $\delta = 0.1$. The minimum of ${\mathcal{L}}^g(\lambda, t)$ appears near the critical point near $\lambda_c = 1-\delta = 0.9$ and around this sharp minimum ${\mathcal{L}}^g(\lambda, t)$ changes rapidly with respect to $\lambda$. As $t$ changes the location of the minimum changes and additional minima also appear. \label{fig1}}
	\end{figure}

	Before investigating the Fisher information it is useful to look closely at the decoherence factor ${\mathcal{L}}^g(\lambda, t)$ which is plotted in Fig.~\ref{fig1} as  function of both $\lambda$ and $t$. We are interested in rapid changes of ${\mathcal{L}}^g(\lambda, t)$ with respect to $\lambda$ because it implies that the non-unitary, $\lambda$-dependent evolution of the state of the probe qubit is quite different for adjacent values of $\lambda$, leading to higher sensitivities in the estimation of $\lambda$. From Fig.~\ref{fig1} we see that graph of ${\mathcal{L}}^g(\lambda, t)$ as a function of $\lambda$ has higher slopes, in magnitude, in the vicinity of the critical point at $\lambda_{c}  = 1 - \delta$. We also note that the nature of ${\mathcal{L}}^g(\lambda, t)$ versus $\lambda$ graph changes as a function of $t$ in a rather involved manner due to the multiple frequencies in Eq.~(\ref{eq:lecho1a}).  
	
\begin{figure}[!htb]
		\resizebox{8.5 cm}{6.2 cm}{\includegraphics{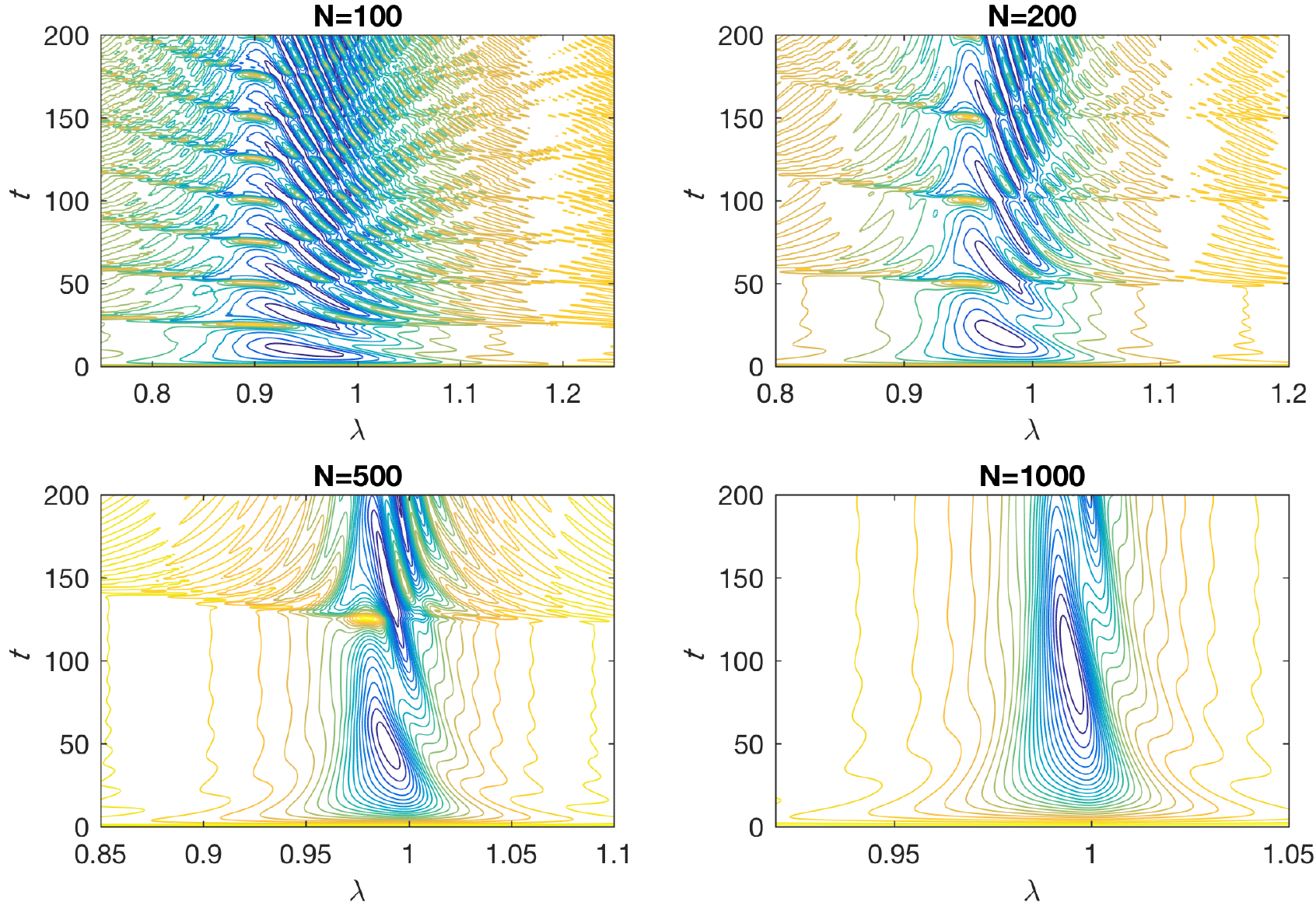}}
		\caption{Contours of the decoherence factor ${\mathcal{L}}^g(\lambda, t)$  plotted against $t$ and $\lambda$ for $N = 100$, $200$, $500$ and $1000$ keeping $N\delta = 10$. Blue contours are for lower values of ${\mathcal{L}}^g(\lambda, t)$ while yellow ones are for higher values. The decoherence factor shows an approximate periodicity  in time as well as a scaling symmetry with respect to the transformation $N \rightarrow N/\gamma$ and $t \rightarrow t/\gamma$ keeping $N/t$ constant. \label{fig1a}}
	\end{figure}

	The contour plots in Fig.~\ref{fig1a} show ${\mathcal{L}}^g(\lambda, t)$ as a function of $\lambda$ and $t$ for several values of $N$. These plots reveal an approximate periodicity for ${\mathcal{L}}^g(\lambda, t)$ as a function of $t$ that we will use further to explore the behaviour of the QFI in this measurement scheme. 
	
\begin{figure}[!htb]
		\resizebox{8.5 cm}{5.5 cm}{\includegraphics{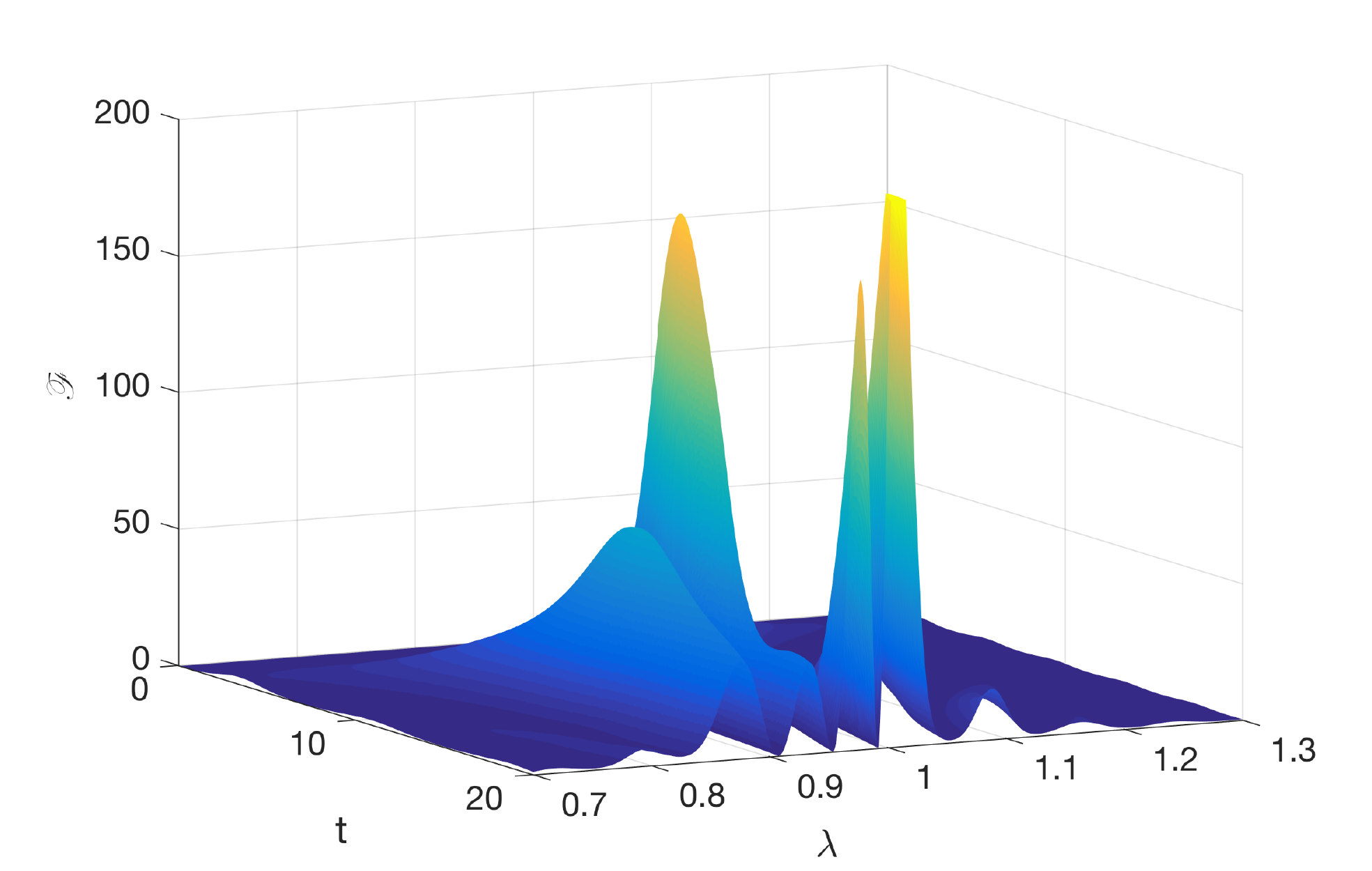}}
		\caption{The QFI ${\mathcal F}^{g}(\lambda,t)$ is plotted against $t$ and $\lambda$ for $N = 100$ and $\delta = 0.1$. There are multiple peaks but the larger maxima of ${\mathcal F}^{g}(\lambda,t)$ appears near the the critical point at $\lambda_c = 1-\delta = 0.9$ and for particular values of $t$. \label{fig2}}
	\end{figure}

	In Fig.~\ref{fig2} we plot the QFI as a function of $\lambda$ and $t$. We see that the QFI has a peak near the effective critical value $\lambda_{c} = 1-\delta$. The peaks on the right of $\lambda_{c}$ are higher than the ones on the left. The behaviour of the Ising ring around $\lambda_{c}$ is dictated by the two competing forces, namely the spin-spin interactions within the ring and the external magnetic field. Below the critical value of the field, the spin-spin interaction wins over the field leading to a phase wherein the Ising spins are aligned along the positive or negative $z$-axis. For $\lambda > \lambda_{c}$ the external field forces the Ising spins to orient along the $x$-axis. Since the interaction between the probe qubit and ring is through the $\sigma_{x}$ operator on the Ising spins, across the critical value of $\lambda$, there is a significant change in the behaviour of decoherence induced on the probe qubit due to the ring which we take advantage of in our measurement scheme.  The QFI having larger peaks on the right side of $\lambda_{c}$ as seen in Fig.~\ref{fig2} is because of the $\sigma_{x}$ coupling to the ring. 
	
	From Fig.~\ref{fig2} we see that the value of QFI has an involved $t$ dependence as well. While $\lambda_{c}$ provides a good choice for the operating point around which this measurement scheme may be implemented, the time dependence of the QFI information raises the question as to whether there is a good operating point in time that one can choose.  Time becomes a free resource if the external magnetic field strength $\lambda$ that we are estimating remains constant. In this case for each value of $N$ we are free to seek out an optimal operating point in the $\lambda - t$ plane wherein the QFI is maximal, and the measurement sensitivity is very good. From Fig.~\ref{fig1a}  we see that the point where the contours of ${\cal L}$ tighten along the $\lambda$-axis appear at different values of $t$ for each value of $N$ and that the first of these points progressively moves away from $t=0$ as $N$ increases. If the $\lambda$ to be estimated is indeed a constant, then all the peaks of the QFI are available for use as an operating point for all $N$. However, if $\lambda$ has a typical time scale in which it changes substantially, then time is also a constrained resource. In that case, one may be forced to choose the first peak of the QFI as a function of $t$ as the operating point, and even this can fail when $N$ becomes larger. 
	
\begin{figure}[!htb]
		\resizebox{8.5 cm}{5.5 cm}{\includegraphics{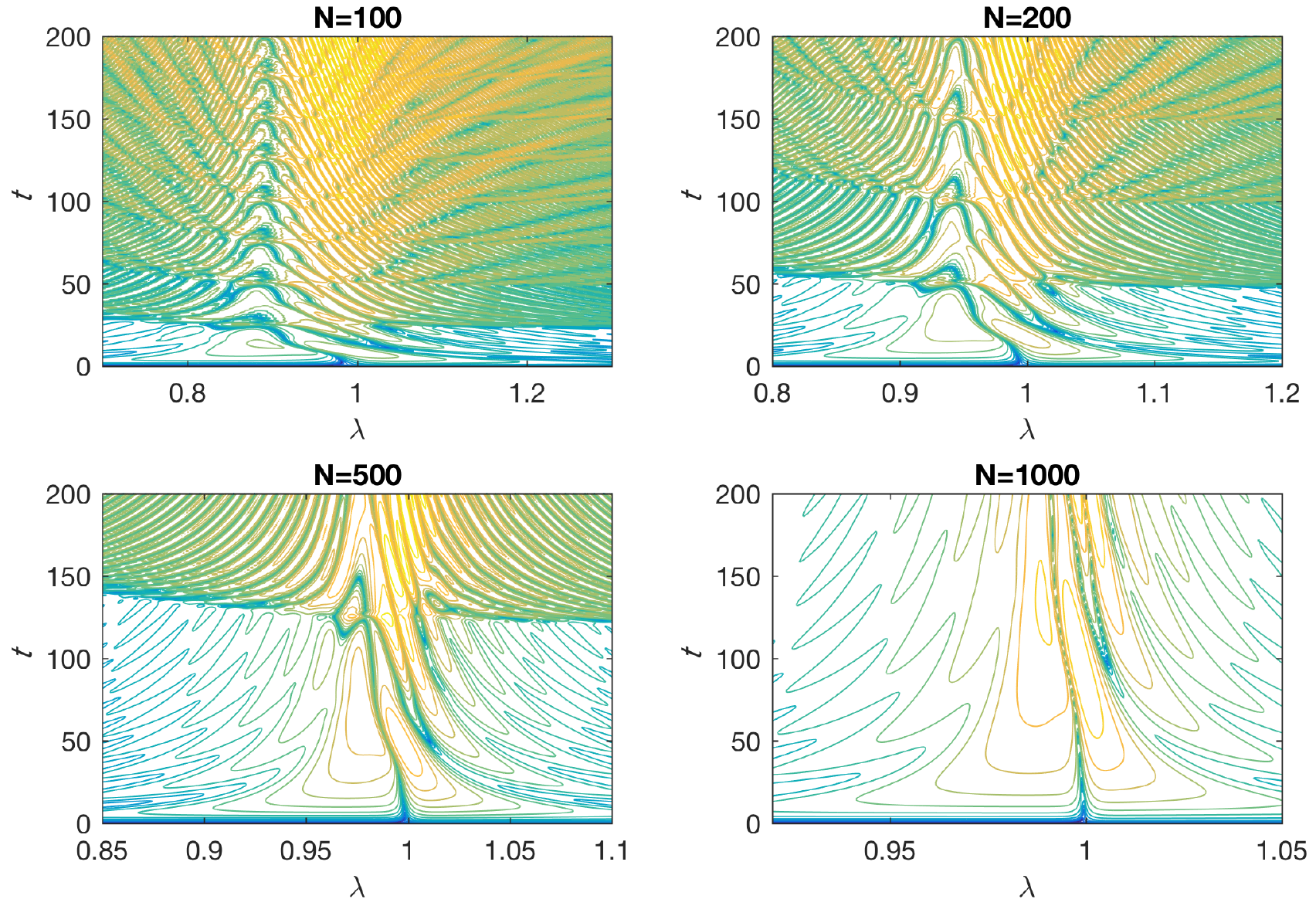}}
		\caption{The contours show the logarithm of the QFI, $ \log_2 ({\mathcal F}^{g}(\lambda, t))$ as a function of $\lambda$ and $t$ for $N = 100$, $200$, $500$ and $1000$ while keeping $N\delta = 10$. The QFI shows a periodicity in $t$ with the period depending on $N$. \label{fig1b}}
	\end{figure}

	The periodicity of the QFI as a function of $t$ can be seen from Fig.~\ref{fig1b} where contour plots of the logarithm of the QFI as a function of $\lambda$ and $t$ are shown. The observed periodicity of the QFI is the same as that of ${\mathcal L}^g$. For the contour plots in both Figs.~\ref{fig1a} and \ref{fig1b} we have kept the product $N\delta$ a constant. This ensures that the interaction energy between the probe qubit and the Ising ring given by the third term in the Hamiltonian in Eq.~(\ref{eq:hamil}) is the same for all $N$. Enhancements in the QFI and measurement precision due to the parameter dependent open evolution arising only due to the increased interaction energy between the Ising ring and the probe is avoided by this choice and we can focus on the enhancement in precision due to changes in the nature of the open evolution of the probe with increasing number of spins in the ring.  
	
\begin{figure}[!htb]
		\resizebox{8.5 cm}{5.5 cm}{\includegraphics{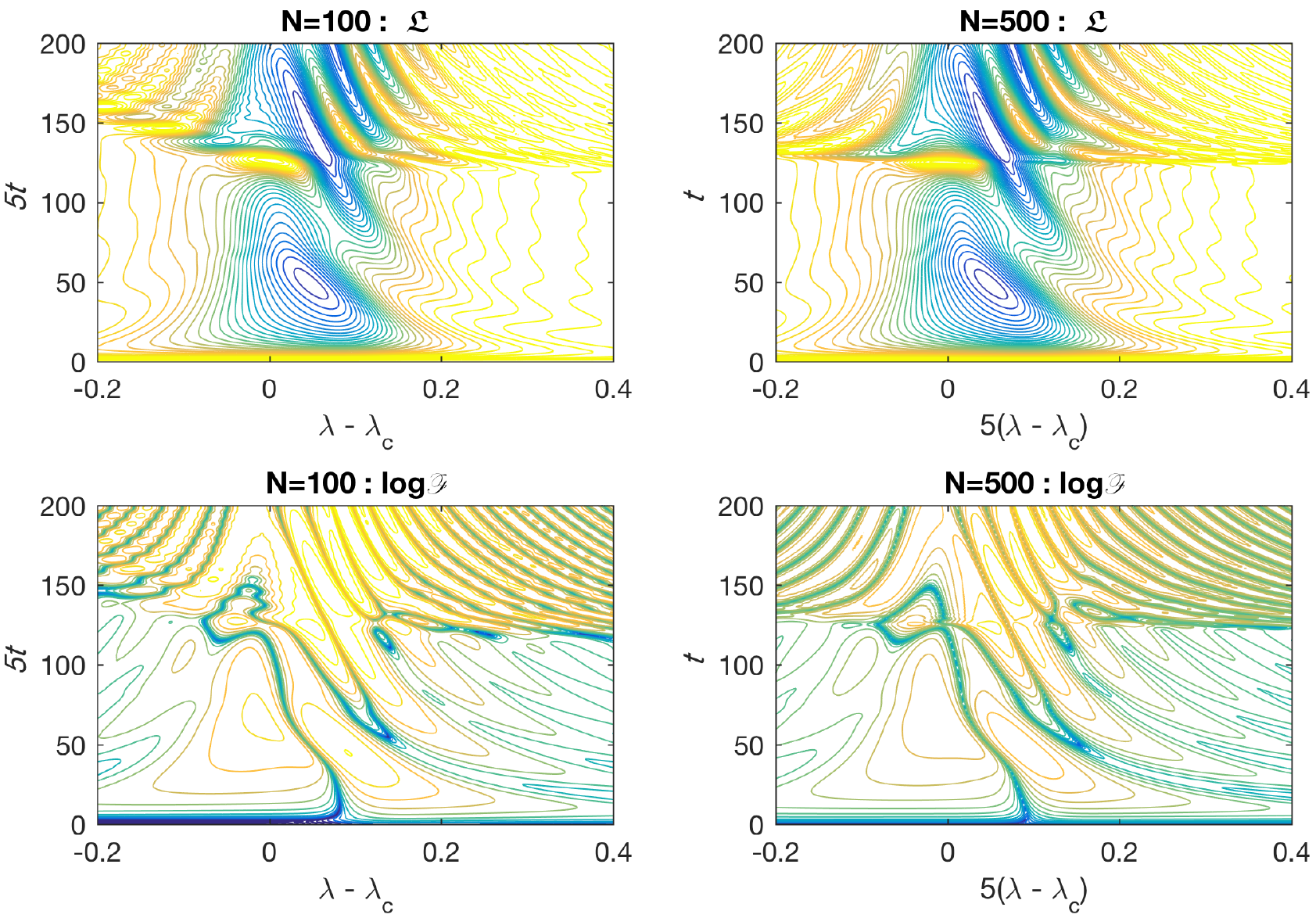}}
		\caption{The contours show the decoherence factor ${\cal L}$ as well as the logarithm of the QFI, $ \ln ({\cal F}^{g})$ as a functions of $\lambda - \lambda_{c}$ and $t$ for $N = 100$ and $500$ keeping $N\delta = 10$. The axes of the plots have been scaled as in Eqs.~(\ref{eq:Lsymm}) and (\ref{eq:Fsymm}) revealing the approximate symmetries of the functions under the scaling through the obvious similarities between the two plots. \label{fig1c}}
	\end{figure}

\begin{figure}[!htb]
		\resizebox{8.5 cm}{3.3 cm}{\includegraphics{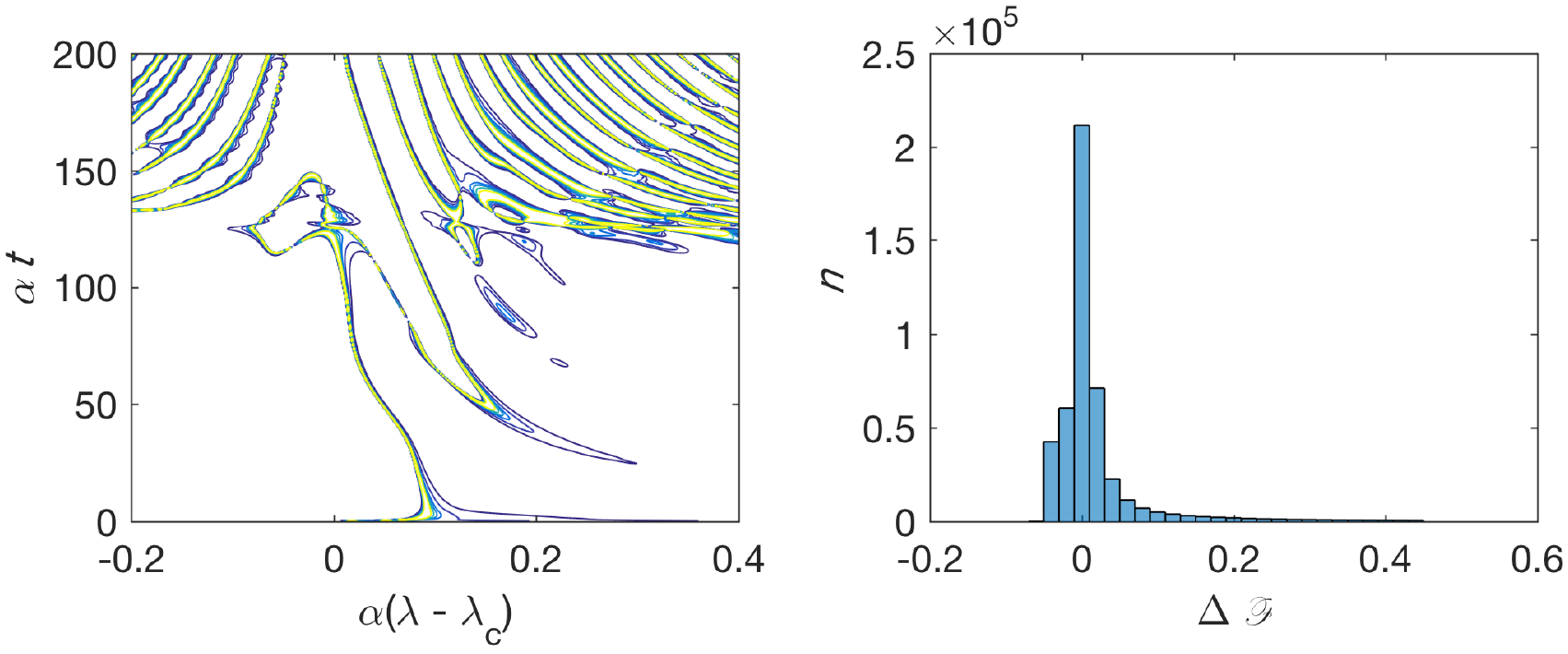}}
		\caption{The plot on the left show contours of $\Delta {\cal F}^g$ as a function of $\lambda - \lambda_{c}$ and $t$. At most of the points $\Delta {\cal F}$ is zero indicating numerically that the symmetry in Eq.~(\ref{eq:Fsymm}) holds. The blue (outermost in each group) contours indicate where $\Delta {\mathcal F}^g$ is zero. All the white regions outside the outermost contour are places where $\Delta {\mathcal F}^g =0$. Only at the places where ${\mathcal F}^g$ is changing rapidly, is $\Delta {\mathcal F}^g$ non-zero. These non-zero values are partly attributable to the finite resolution in the $\lambda$ - $t$ plane of the numerical computations. The histogram on the right shows the distribution of $\Delta {\cal F}^g$ on the nearly 500,000 points of the $\lambda - t$ grid on which $\Delta {\cal F}^g$ was computed for the plot on the left. The histogram again shows that on almost all the grid points, $\Delta {\cal F}^g$ is zero. \label{fig1d}}
	\end{figure}

	The contour plots in Figs.~\ref{fig1a} and \ref{fig1b} also suggest scaling symmetries for both ${\mathcal{L}}^g(\lambda, t)$ and ${\mathcal F}^{g}(\lambda,t)$. Choosing to re-parametrise both ${\mathcal{L}}^g(\lambda, t)$ and ${\mathcal F}^{g}(\lambda,t)$ in terms $\lambda - \lambda_{c}$ in place of $\lambda$, numerically we find an approximate symmetry for ${\mathcal{L}}^g(\lambda, t)$ wherein 
	\begin{equation}
	\label{eq:Lsymm}
	{\cal L}^g ((\lambda - \lambda_{c})/\alpha, \alpha t, \alpha N, \delta/\alpha) \simeq {\cal L}^g(\lambda-\lambda_{c}, t, N ,\delta). 
	\end{equation}
	From Eq.~(\ref{Fqubit}), it follows that
	\begin{equation}
	\label{eq:Fsymm}
	{\cal F}^g ((\lambda - \lambda_{c})/\alpha, \alpha t, \alpha N, \delta/\alpha) \simeq \alpha^{2}{\cal F}^g(\lambda-\lambda_{c}, t, N ,\delta). 
	\end{equation}
	This approximate symmetry can be seen by inspection in Fig.~\ref{fig1c} where contours of both ${\cal L}^g$ and ${\ln({\cal F}^g)}$ for $N=100$ and $N=500$ respectively are plotted as functions of $\lambda-\lambda_{c}$ and $t$ with $\lambda_{c} = 1-\delta$ and $N\delta = 10$. 	
	In the plots on the left for $N=100$, the time axis is stretched by a factor of 5 while in the plots on the right for $N=500$, $\lambda-\lambda_{c}$ is scaled up by the same factor in accordance with the numerically uncovered symmetries for ${\cal L}^g$ and ${\cal F}^g  $ in Eqs.~(\ref{eq:Lsymm}) and (\ref{eq:Fsymm}). The obvious similarities between the graphs on the right and those on the left are quantified in Fig.~\ref{fig1d} where on the left the contours of 
	\[ \Delta {\cal F}^g \equiv \frac{{\cal F}^g_{100}( \lambda-\lambda_{c}, t) - {\cal F}^g_{500}( (\lambda - \lambda_{c})/\alpha,\alpha t)/\alpha^{2}}{{\cal F}^g_{500}} \]
	are plotted against $t$ and $\lambda - \lambda_{c}$. We see that $\Delta {\cal F}^g$ is zero for almost all values of $\lambda$ and $t$ after the axes have been suitably scaled for ${\cal F}^g_{500}$ as per the scaling invariance in Eq.~(\ref{eq:Fsymm}).
	
\begin{figure}[!htb]
		\resizebox{8 cm}{5.5 cm}{\includegraphics{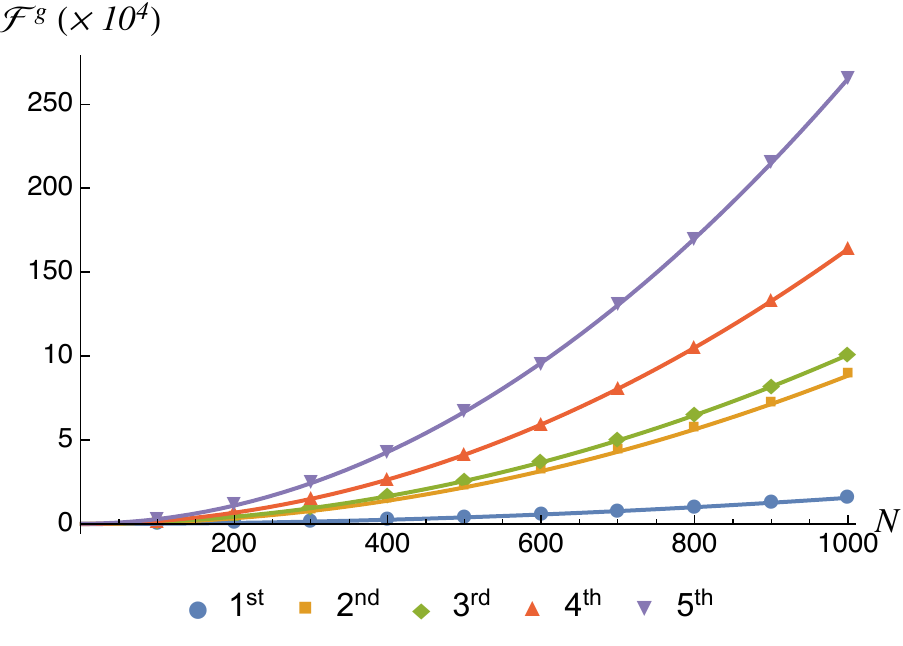}}
		\caption{ $\mathcal{F}^{g}$ is plotted as a function of $N$ for the first five peaks of the QFI in the $\lambda$-$t$ plane. The continuous lines show the quadratic fit functions for the increase in height of each peak with increasing $N$. The quadratic increase in ${\cal F}^g$ with respect to $N$ means that the measurement uncertainty $\delta \lambda$ exhibits the Heisenberg limited scaling of $1/N$.   \label{fig:g}}
	\end{figure}

	In the parameter regime where the numerically revealed approximate symmetry for the QFI in Eq.~(\ref{eq:Fsymm}) holds, we expect Heisenberg limited scaling for the measurement uncertainty since 
	\[ {\cal F}^g(N=\alpha N_{0}) = \alpha^{2} {\cal F}^g(N_{0})\]
	and taking $N_{0}=1$ we see that 
	\[ \delta \lambda \propto \frac{1}{\sqrt{\cal F}^g} = \frac{1}{N}. \]
	To numerically verify this expected scaling, we computed ${\cal F}^g$ corresponding to the first five peaks in the $\lambda$ - $t$ plane for increasing $N$. These values are shown in  Fig.~\ref{fig:g}.  For all the cases we choose the product $N\delta = 10$. Numerical fits for the $\mathcal{F}^g$ to quadratic functions of the form 
	\[ {\cal F}^{g}_{N} = a_{l} N^{2} + b_{l} N + c_{l},\]
	where `$l$' labels the peak of ${\mathcal F}^g$ that is followed as a function of $N$, are also shown in the figure as continuous lines. We see from Fig.~\ref{fig:g} that ${\cal F}^{g}$ does scale as $N^{2}$ leading to Heisenberg limited scaling for the measurement precision in the metrology scheme we consider. 	

	\section{Ising Chain Initialized in the Thermal State \label{thermalFid}}

	Putting the Ising ring in its ground state is a challenging task that may not be achievable using available technologies within the constraints of a practical measurement scheme. In this section we study the QFI of the probe qubit for the case where the Ising ring is in a thermal state,
	\[ \rho_{\rm in}=\frac{e^{-\beta H_0}}{{\rm tr}(e^{-\beta H_0})}, \]
	where $\beta=1/T$ and $T$ is the temperature. As shown in Appendix~\ref{sec:ThermalSubSection}, the relevant decoherence factor in this case is obtained as
	\begin{equation}
	\label{eq:thermalLE}
	\mathcal{L}^T=\big|{\rm tr} [e^{iH_1 t} e^{-iH_0t} \rho_{\rm in}]\big|^2=  \prod_k  \frac{p_k^2+q_k^2}{\big[1+\cosh (\beta \epsilon_k^0) \big]^2},
	\end{equation}
	where 
	\begin{eqnarray*}
		p_k & = & \cosh(\beta \epsilon_k^0)\big[ \cos(\epsilon_k^1 t) \cos(\epsilon_k^0 t) \\
		&& \quad + \, \sin(\epsilon_k^1 t) \sin(\epsilon_k^0 t) \cos(2 \alpha_k) \big] + 1
	\end{eqnarray*}
	\begin{eqnarray*}
		q_k & = & \sinh(\beta \epsilon_k^0)\big[ \cos(\epsilon_k^1 t) \sin(\epsilon_k^0 t) \\
		&& \quad - \, \sin(\epsilon_k^1 t) \cos(\epsilon_k^0 t) \cos(2 \alpha_k) \big],
	\end{eqnarray*} 
	with $\alpha_k$, $\epsilon_k^0$, and $\epsilon_k^1$ having the same definitions as before. Starting from the decoherence factor we can again compute the QFI, ${\mathcal F}^T$ for the probe spin (See Appendix~\ref{sec:ThermalSubSection}). 
	\begin{figure}[!htb]
		\resizebox{8.5 cm}{6 cm}{\includegraphics{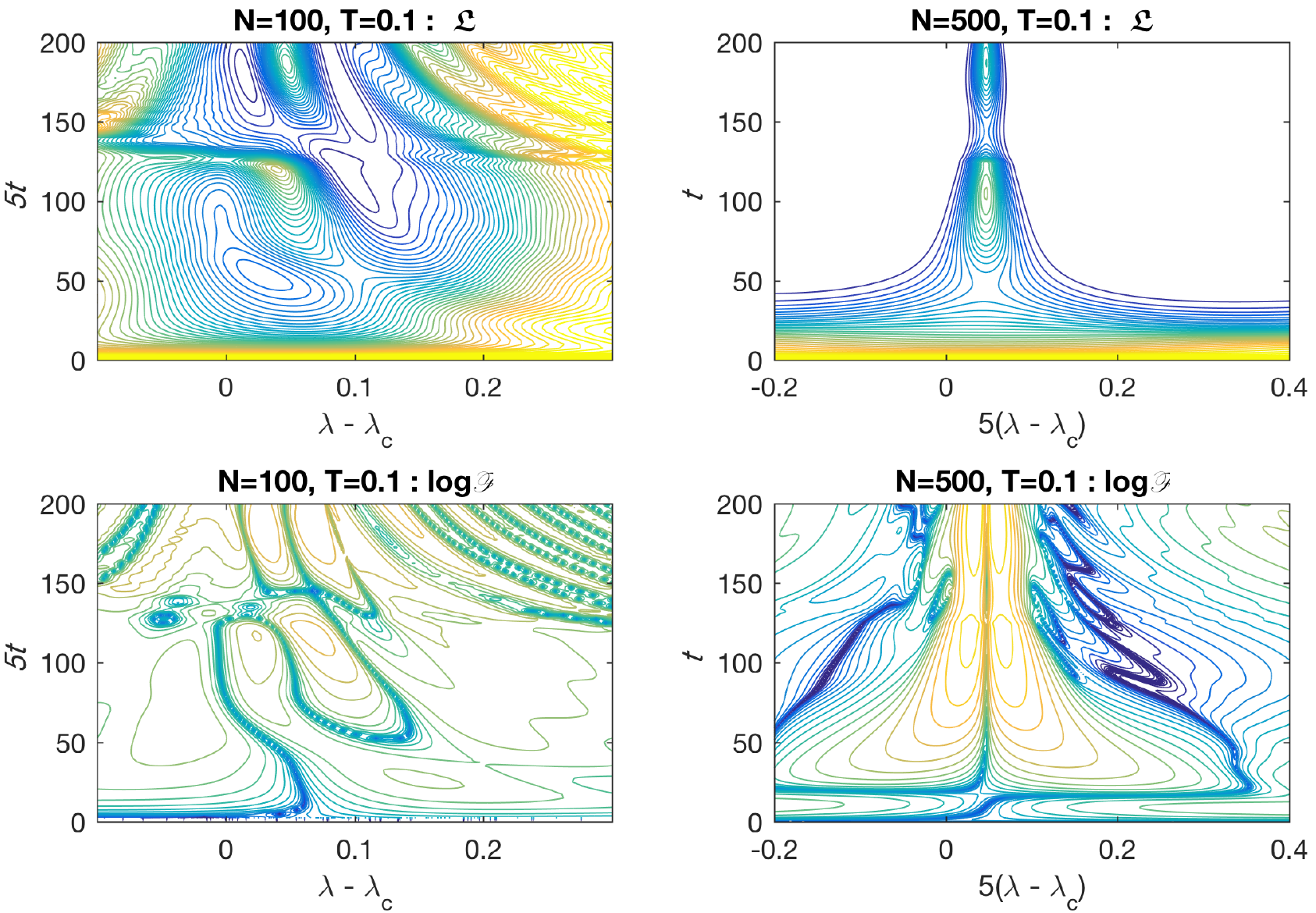}}
		\caption{The contours show the decoherence factor ${\cal L}^T$ as well as the logarithm of the QFI, $ \ln ({\cal F}^{T})$ as a functions of $\lambda - \lambda_{c}$ and $t$ for $N = 100$ and $500$ keeping $N\delta = 10$ and $T=0.1$. The axes of the plots have been scaled as in Eqs.~(\ref{eq:Lsymm}) and (\ref{eq:Fsymm}). The scaling symmetry and periodicity of both ${\mathcal L}^T$ and ${\mathcal F}^T$ can be still be seen in the contour plots however, especially in the case of the QFI, the function is flattened and the number of contours reduced by the effect of the finite temperature. The effect of the temperature is more pronounced in the case where $N$ is larger. \label{figThermalContour1}}
	\end{figure}
	
	The approximate periodicity of ${\mathcal L}^g(\lambda,t)$ and ${\mathcal F}^g(\lambda, t)$ that was observed numerically is found to applicable to ${\mathcal L}^T$ and ${\mathcal F}^T$ as well in the case where the Ising ring is in a thermal state. This can be seen from Fig.~\ref{figThermalContour1}, where contours of the decoherence factor as well as those of $\ln ({\mathcal F}^T)$, corresponding to $N=100$ and $500$ for the thermal state case are plotted for a finite temperature and with suitably scaled axes. We see that in this case while the periods of the function match under the scaling, the similarities between the two sets of contours is not as evident as in the zero temperature case discussed in the previous section. There are two aspects to be considered here. One is the effect of the finite temperature and the other is the unequal ways in which the finite temperature affects the $N=100$ and $N=500$ cases for the $t$ and $\lambda$ values considered. The effect of the finite temperature can be explored by looking at the asymptotic behavior of ${\mathcal L}^T(\lambda, t)$ as a function of the temperature. When $T \rightarrow 0$ ($\beta \rightarrow \infty$), ${\mathcal L}^T \rightarrow {\mathcal L}^g$ as expected. On the other hand when $T \rightarrow \infty$ ($\beta \rightarrow 0$), ${\mathcal L}^T \rightarrow 0$ which means that at large temperatures ${\mathcal F}^T = 0$. So we cannot expect the scaling symmetry in Eqs.~(\ref{eq:Lsymm}) and (\ref{eq:Fsymm}) to be applicable to the thermal state case even if the periodicity of the two sets of functions are the same. By comparing Figs.~\ref{fig1c} and \ref{figThermalContour1} we see that for larger $N$, the temperature affects the decoherence factor and the QFI to a larger degree for all values of $\lambda$ and $t$. The contours for the $N=100$ case are similar for $T=0$ and $T=0.1$ in units where $\hbar = k_B = 1$, while the corresponding contours for $N=500$ show substantial differences.

\begin{figure}[!htb]
		\centering	
		\resizebox{8 cm}{5.1 cm}{\includegraphics{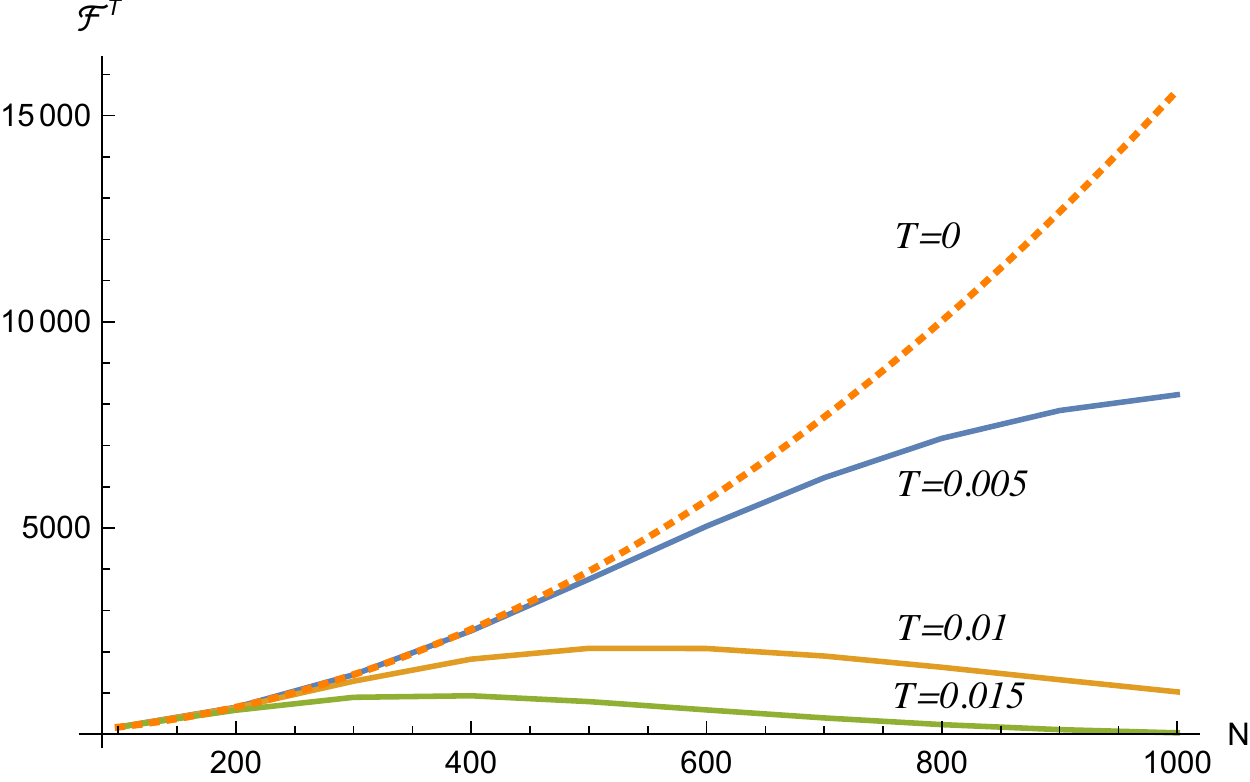}}
		\caption{ $\mathcal{F}^{T}$ is plotted as a function of $N$ for the first peak of the QFI in the $\lambda$-$t$ plane for temperatures $T = 0,\;0.005,\;0.01,\;0.015 $ in units where $\hbar = k_B = 1$, keeping $N \delta=10$. We see that at finite temperatures and large $N$, not only is the quadratic behavior of ${\mathcal F}^T$ lost, but the QFI can even go below the shot noise limit (linear growth) levels. }
		
		\label{fig:Thermal}
	\end{figure}

	The effect of temperature on the scaling of the QFI, $\mathcal{F}^T$ with $N$ is seen in Fig.~\ref{fig:Thermal}. Using the periodicity of the functions, we plot $\mathcal{F}^T$  versus $N$ only for the first peak of the QFI in the $\lambda$-$t$ plane for temperatures $T = 0,\;0.005,\;0.01,\;0.015$ keeping $N\delta=10$. As expected at $T=0$, ${\mathcal F}^T = {\mathcal F}^g$ Fig.~\ref{fig:Thermal} also shows that for $T \neq 0$,  $\mathcal{F}^T$ deviates from the $\mathcal{F}^g$ for large $N$ values and the range of $N$ through which ${\cal F}^T$ scales as $N^2$ reduces as the temperature of the Ising ring increases. This means that Heisenberg-limited scaling for the measurement uncertainty is available only through a finite range in $N$ depending on the temperature.

	\section{Discussion and Conclusion \label{conclude}}	
	
	Starting from the intuition that around a quantum phase transition that is dependent on an external parameter like an applied magnetic field, the state of the quantum system undergoing the transition will change rapidly with respect to the parameter, we explored the possibility of constructing a precision magnetometer working with Heisenberg-limited sensitivities using a probe qubit coupled to an Ising ring of spins operating near a critical point. We found that in the zero temperature case, Heisenberg-limited scaling for the measurement uncertainty in the value of the applied transverse magnetic field is indeed possible. In fact, given the sudden and rapid change of state of a system across a critical point, it is reasonable to expect a performance that is better than the Heisenberg-limited scaling of $1/N$. However we find that this is not the case, and the quantum Cramer-Rao bound prevails even when phase transitions are involved because the measured parameter is coupled linearly and independently to the individual probe units (Ising spins)~\cite{Boixo:2007ik,Boixo:2008bc}.
	
	Our analysis of the Ising ring coupled to the probe qubit shows that not only is it advantageous to work around a quantum phase transition for obtaining enhanced precision in a quantum limited measurement, but also that non-unitary evolution of the probe can also contribute to the quantum advantage in metrology. We were able to study the contribution to the measurement precision from this non-unitary part of the time evolution of the probe in isolation by choosing the free evolution term in the Hamiltonian for the probe qubit to be zero. The Heisenberg-limited scaling we obtain for the zero temperature case is therefore entirely from the non-unitary part of the open evolution of the probe in contact with the Ising ring which constitutes its immediate environment. Typically, open evolution and decoherence are considered as challenges to be overcome in implementing quantum technologies. Here we show that such evolution can be leveraged to give quantum advantages in the context of quantum limited measurements. 
	
	Precision magnetometry, as mentioned previously, is a field of significant applied interest. We have analyzed an idealized model that provides Heisenberg-limited scaling for the measurement precision for magnetic field sensing. In our scheme there is no need to initialize a large number of qubits in particular entangled states to obtain the scaling. To have the $1/N$ scaling for arbitrary ranges of $N$ we require the Ising ring to be in the ground state ($T=0$) however. If the scaling is needed only for shorter ranges of $N$ then thermal states at low enough temperatures are seen to be sufficient. Treating the Ising ring as the environment and not part of the quantum probe can also be called into question because to obtain the Heisenberg limited scaling, even the ring will have to be engineered in order to initialize it in the ground state. Extending our analysis to more complex models that better represent the type of systems that can be engineered, or occurs naturally, in the laboratory remains to be done as future work. 
	
	\acknowledgements
	
	This work is supported in part by a grant from SERB, DST, Government of India No. EMR/2016/007221 and by the Ministry of Human Resources Development, Government of India through the FAST program. N. J. acknowledges the support of the IMPRINT program of the Government of India through grant no.~6099.

	\bibliography{FisherCentral} 
	
	\appendix
	\section{Computing the Decoherence factor \label{App_thermalLE}}
	
	To obtain an expression for $\mathcal{L}^g(\lambda, t)$  and $\mathcal{L}^T(\lambda, t)$ we need to first diagonalise the Hamiltonians $H_1$ and $H_0$. As shown in~\cite{sachdev2007quantum} a Hamiltonian of the form 
	\[ H_\gamma=-\sum_j (\sigma_j^z \sigma_{j+1}^z +h_\gamma \sigma_j^x), \] 
	can be mapped on to the quasi-free Fermionic Hamiltonian of the form 
	\begin{eqnarray}
	\label{eq:Hgamma}
	H_\gamma & = & \sum_{k} H_{\gamma k} \nonumber \\
	& = &  -\sum_k \big\{\left[\cos (k)- h_\gamma\right] (c_k^\dagger c_k^{}  + c_{-k}^\dagger c_{-k}^{})\nonumber \\
	&& \qquad + \, i\sin(k)\; (c_{-k}^\dagger c_k^\dagger +c_{-k}^{} c_k^{})\big\}, 
	\end{eqnarray}
	where  $\gamma=0,1$ with $h_0 = \lambda$, $h_1 = \lambda + \delta$ and the $c_k$'s are anticommuting Fermionic operators. The mapping is done by first taking Jordan-Wigner transformation~\cite{jordanwigner28,sachdev2007quantum}  to map the $\sigma^{x,y,z}_j$ to their respective Fermionic operators and then taking the Fourier transformation replacing the position space Fermionic operators with the corresponding momentum space ones. Here we are assuming unit spacing between the spins in the Ising ring.
	
	Each of the $H_{\gamma k}$ can now be diagonalized using a Bogoliubov transformation of the form
	\begin{equation}
	\label{eq:dlambda}	
	d_{\gamma, k}= \cos(\theta^\gamma_k/2) c_k - i \sin (\theta^\gamma_k/2) c_{-k}^\dagger,
	\end{equation}
	where
	\begin{equation}
	\theta_k^\gamma=\tan^{-1}\left[\frac{\sin(k)}{h_\gamma - \cos (k)}  \right].
	\label{eq:theta}
	\end{equation} 
	We then obtain, 
	\begin{equation}
	\label{eq:Hdiagonal}
	H_\gamma = \sum_k \epsilon^\gamma_k( d_{\gamma, k}^\dagger d_{\gamma, k} - 1/2),
	\end{equation}
	where
	\begin{equation}
	\epsilon_k^\gamma  = 2\sqrt{1+h_\gamma^2-2h_\gamma \cos(k)},
	\label{eq:eps}
	\end{equation}
	
	In the momentum space representation, we see that the Hilbert space of the system factorizes as ${\mathcal H} = \otimes {\mathcal H}_k$ and each of the operators $H_{\gamma k}$ in Eq.~(\ref{eq:Hgamma}) act on distinct subspaces~\cite{QSL+06} given by ${\mathcal H}_k \otimes {\mathcal H}_{-k}$ for each $k$. Since $\epsilon_k^\gamma > 0$ for all $k$, the ground state $|G^\gamma \rangle$ of $H_\gamma$ should be such that $d_{k,\gamma}|G^\gamma \rangle = 0$ for all $k$. Each of the subspaces is spanned by the four vectors, $\{ |00\rangle_{k,-k}, |10\rangle_{k,-k}, |01\rangle_{k,-k}, |11\rangle_{k,-k}\}$ with the 1 and 0 denoting the occupancy of the  momentum states labelled by $k$ and $-k$ respectively. In this basis, using Eq.~(\ref{eq:dlambda}) it is easy to see that 
	\begin{equation}
	\label{eq:groundstates}
	|G^\gamma\rangle = \otimes_k |G^\gamma_k\rangle = \otimes_k \big[  \sin \frac{\theta^\gamma_k}{2} |11\rangle - i \cos \frac{\theta^\gamma_k}{2} |00\rangle \big]. 
	\end{equation}
	Since $|\phi(0)\rangle$ is the ground state of $H_0$, we have $|\phi(0)\rangle = |G^0\rangle$. We therefore have to evaluate
	\[ {\mathcal L}^g = |\langle G^0|e^{iH_0t} e^{-iH_1t} |G^0\rangle|^2. \]
	
	In order to obtain a suitable form for $H^\gamma$ which will allow us to exponentiate the Hamiltonian easily, we note that within each subspace spanned by the four vectors $\{ |00\rangle_{k,-k}, |10\rangle_{k,-k}, |01\rangle_{k,-k}, |11\rangle_{k,-k}\}$, the action of $H_{\gamma k}$ is nontrivial only on the even parity sector spanned by $\{ |00\rangle_{k,-k}, |11\rangle_{k,-k}\}$. Within this two dimensional sector, it is worthwhile to go back to representing $H_\gamma$ in terms of Pauli operators rather than Fermionic ones by using the transformation, 
	\begin{eqnarray*}
		\sigma_{kz} & = & c_k^\dagger c_k^{} + c_{-k}^\dagger c_{-k}^{} -1\\
		\sigma_{ky} & = & -i c_k^\dagger c_{-k}^\dagger + i c_{-k} c_k \\
		\sigma_{kx} & = & c_k^\dagger c_{-k}^\dagger +  c_{-k} c_k.
	\end{eqnarray*}
	Apart from constants which we ignore since they have no bearing on ${\mathcal L}^g$ we have, 
	\begin{equation}
	\label{eq:Hpauli2}
	H_{\gamma k} =  \{ [h_\gamma - \cos(k)]\sigma_{kz} + \sin(k) \sigma_{ky} \} \oplus \openone_{\rm odd}, 
	\end{equation}
	where 
	\[ \openone_{\rm odd} = \arrowvert 1 0 \rangle_{k,-k}\langle 1 0 \arrowvert_{k,-k} + \arrowvert 0 1 \rangle_{k,-k}\langle 0 1 \arrowvert_{k,-k}, \] 
	represents the trivial action of $H_{\gamma k}$ in the odd parity sector. We will not explicitly indicate the action on the odd parity sector henceforth because it again has no effect on ${\mathcal L}^g$ as $|G^\gamma_k\rangle$ have no support on the odd parity sectors. Observing that $H_{\gamma k}$ can be obtained from $\sigma_{kz}$ by a rotation about the $\sigma_{kx}$ axis followed by a scaling, we can write $H_{\gamma k}$ in an alternate diagonal form as, 
	\[ H_{\gamma k}  = e^{i (\theta^\gamma_k/2)\sigma_{kx} } \epsilon^\gamma_k \sigma_{kz} e^{-i (\theta^\gamma_k/2)\sigma_{kx} }, \]
	and it follows that for any $\beta$,
	\begin{equation}
	\label{eq:Hexp}
	e^{\beta H_\gamma} = e^{\beta \sum_k H_{\gamma k}} = \prod_k  e^{i \frac{\theta^\gamma_k}{2}\sigma_{kx} } e^{\beta \epsilon^\gamma_k \sigma_{kz}} e^{-i \frac{\theta^\gamma_k}{2}\sigma_{kx} }.
	\end{equation}
	So we have
	\begin{eqnarray*}
		{\mathcal L}^g & = & \prod_k |\langle G^0_k|e^{iH_{0k}t} e^{-iH_{1k}t} |G^0_k\rangle|^2 \\
		& =& \prod_k |\langle G^0_k| e^{i (\theta^0_k/2)\sigma_{kx} } e^{i\epsilon^0_k \sigma_{kz}} e^{-i \alpha_k \sigma_{kx} } \\
		&& \quad \times \, e^{i (\theta^\gamma_k/2)\sigma_{kx} } e^{-i \epsilon^1_k \sigma_{kz}} e^{-i (\theta^1_k/2)\sigma_{kx} } |G^0_k\rangle|^2,
	\end{eqnarray*}
	where $\alpha_k = (\theta_k^0 - \theta_k^1)/2$. Expanding each of the exponentials in Euler form and using $\sigma_x|00\rangle = |11\rangle$, $\sigma_x |11\rangle = |00\rangle$, $\sigma_z |00\rangle = -|00\rangle$ and $|\sigma_z|11\rangle = |11\rangle$, we obtain,
	\begin{equation}
	\mathcal{L}^g(\lambda, t)=\prod_{k}\left[1-\sin^2(2\alpha_k)\sin^2(\epsilon_k^1 t)\right],
	\label{eq:lecho1}
	\end{equation}
	with $k=2n\pi/N$ and $n=1,2, \ldots, N/2$.

	\subsection{Thermal state \label{sec:ThermalSubSection}}
	
	The initial state of the probe qubit and the Ising ring is assumed to be a product state of the form
	\[ \rho_{P\!R} = \eta_P \otimes \rho_{\rm in}, \quad \rho_{\rm in} = \frac{e^{-\beta H_0}}{{\rm tr}(e^{-\beta H_0})}.\]
	where 
	\[ \eta_{\rm P} = \eta_{00}|0\rangle \langle 0| + \eta_{01} |0\rangle \langle 1| + \eta_{10} |1 \rangle \langle 0| + \eta_{11} |1\rangle \langle 1|. \]
	The joint evolution of the probe qubit and the ring generated by the Hamiltonian in Eq.~(\ref{eq:hamil}) leads to the time evolved state
	\begin{eqnarray*}
		\tilde{\rho}_{\rm P R} & = & \eta_{00} |0\rangle \langle 0| \otimes e^{-iH_0t} \rho_{\rm in} e^{iH_0t} \\
		&& \quad + \,  \eta_{01} |0\rangle \langle 1| \otimes e^{-iH_0t} \rho_{\rm in} e^{iH_1t} \\ 
		&& \quad + \, \eta_{10} |1\rangle \langle 0| \otimes e^{-iH_1t} \rho_{\rm in} e^{iH_0t}
		\\ 
		&& \quad + \, \eta_{11} |1\rangle \langle 1| \otimes e^{-iH_1t} \rho_{\rm in} e^{iH_1t}.
	\end{eqnarray*} 
	Using the fact that $ e^{-iH_0t} \rho_{\rm in} e^{iH_0t}$ and $ e^{-iH_1t} \rho_{\rm in} e^{iH_1t}$ are states of the ring with unit trace, we can trace out the Ising ring from $\tilde{\rho}_{P\!R}$ to get
	\begin{eqnarray*}
		\tilde{\eta}_{\rm P} & = &  \eta_{00}|0\rangle \langle 0| + e^{-i \varphi(\delta) t} \sqrt{{\mathcal L}^T}  \eta_{01} |0\rangle \langle 1| \\
		&& \quad + \, e^{i \varphi(\delta) t}  \sqrt{{\mathcal L}^T} \eta_{10} |1 \rangle \langle 0| + \eta_{11} |1\rangle \langle 1|
	\end{eqnarray*}
	where $\varphi(\delta)$ is a phase that depends only on the coupling $\delta$ and hence is not of interest to us while estimating $\lambda$. The decoherence factor ${\mathcal L}^T$ is given by 
	\[ \mathcal{L}^T=\big|{\rm tr} [e^{iH_1 t} e^{-iH_0t} \rho_{\rm in}]\big|^2=\big|{\rm tr} [e^{iH_0 t} e^{-iH_1t} \rho_{\rm in}]\big|^2.\]
	The tensor product, factorized, structure of the Hilbert space of states of the ring in the momentum representation means that
	\[ \rho_{\rm in}  =  \prod_k \frac{1}{{Z_k^0}} \rho_k^0 \oplus \openone_{\rm odd} \]
	Using Eq.~(\ref{eq:Hexp}) we get,	 
	\[ e^{iH_1 t} e^{-iH_0t} \rho_{\rm in} = \prod_k \frac{1}{Z_k^0} e^{iH_{1k} t} e^{-(\beta +it)H_{0k}} \oplus \openone_{\rm odd},\]
	where~\cite{QSL+06},
	\[ Z_k = {\rm tr}(e^{-\beta H_{0k}} \oplus \openone_{\rm odd}) = 2 + 2\cosh( \beta \epsilon^0_k ). \]
	We therefore have
	\[ {\mathcal L}^T = \prod_k \frac{|{\rm tr}( e^{iH_{1k} t} e^{-(\beta +it)H_{0k}})+2|^2}{[2 + 2\cosh( \beta \epsilon^0_k )]^2}. \]
	Expanding the exponentials in Euler form, we get
	\begin{eqnarray*}
		{\rm tr} \big[ e^{iH_{1k} t} e^{-(\beta +it)H_{0k}} \big] & = & 2 \cosh(\beta \epsilon^0_k) [ \cos (\epsilon^0_k t) \cos ( \epsilon^1_k t)\\
		&& \quad \;   + \cos (2 \alpha_k) \sin (\epsilon^0_k t) \sin ( \epsilon^1_k t)] \\
		&& +2 i \sinh(\beta \epsilon^0_k) [ \sin (\epsilon^0_k t) \cos (\epsilon^1_k t) \\
		&& \quad \;   - \cos (2 \alpha_k) \cos ( \epsilon^0_k t) \sin (\epsilon^1_k t)]
	\end{eqnarray*} 
	Using the equation above, we obtain,	
	\begin{equation}
	\label{eq:thermalLE}
	\mathcal{L}^T=  \prod_k  \frac{p_k^2+q_k^2}{\big[1+\cosh (\beta \epsilon_k^0) \big]^2},
	\end{equation}
	where 
	\begin{eqnarray}
	\label{eq:pkqk}
	p_k & = & \cosh(\beta \epsilon_k^0)\big[ \cos(\epsilon_k^1 t) \cos(\epsilon_k^0 t) \nonumber \\
	&& \quad + \, \sin(\epsilon_k^1 t) \sin(\epsilon_k^0 t) \cos(2 \alpha_k) \big] + 1 \nonumber \\
	q_k & = & \sinh(\beta \epsilon_k^0)\big[ \cos(\epsilon_k^1 t) \sin(\epsilon_k^0 t) \nonumber \\
	&& \quad - \, \sin(\epsilon_k^1 t) \cos(\epsilon_k^0 t) \cos(2 \alpha_k) \big].
	\end{eqnarray} 
	The logarithmic derivative of ${\mathcal L}^T$ with respect to $\lambda$ is obtained as
	\[
	\frac{\partial {\mathcal L}^T}{{\mathcal L}^T}  =  2 \sum_k \bigg\{ \frac{p_k \partial_\lambda p_k + q_k \partial_\lambda q_k}{p_k^2 + q_k^2} - \, \beta \frac{\sinh(\beta \epsilon_k^0) \partial_\lambda \epsilon^0_k}{1+\cosh(\beta \epsilon^0_k)} \bigg\}.
	\]
	The derivatives $\partial_\lambda p_k$ and $\partial_\lambda q_k$ can be obtained from Eq.~(\ref{eq:pkqk}) through a straightforward differentiation wherein it is convenient to use
	\[ \cos(2 \alpha_k) = \frac{4[1+\lambda(\lambda+\delta)-(2 \lambda+\delta) \cos(k)]}{\epsilon^0_k \epsilon^1_k}, \]
	and
	\[ \partial_\lambda(\cos(2 \alpha_k)) = \frac{64 \delta^2[2 \lambda +\delta - 2\cos(k)] \sin^2(k)}{(\epsilon^0_k)^2(\epsilon^1_k)^2}. \]

\end{document}